\documentclass[12pt]{article}
\usepackage{color}
\usepackage{amssymb}
\usepackage[pdftex]{graphics}
\usepackage{graphicx}
\usepackage{times}
\usepackage{cite}

\def\ket#1{\mathinner{|{#1}\rangle}}

\begin{document}
\begin{center}
{\bf Topological Insulators with Ultracold Atoms}

Indubala I Satija and Erhai Zhao\\
School of Physics, Astronomy and Computational Sciences, \\
George Mason University,
 Fairfax, VA 22030, USA 
\end{center}

Ultracold atom research presents many avenues to study problems at the forefront of physics.
Due to their unprecedented controllability, these systems are ideally suited to explore
new exotic states of matter, which is one of the key driving elements of the
condensed matter research. One such topic of considerable importance is
topological insulators, materials
that are insulating in the interior but conduct
along the edges. Quantum Hall and its close cousin Quantum Spin Hall states belong to the family of these exotic states
and are the subject of this chapter. 

\section*{Contents}
\begin{itemize}
\item Ultracold atoms, synthetic gauge fields, and topological states  ......... 1 
\item Topological aspects: Chern numbers and edge states...........................3
\item The Hofstadter model and the Chern number........................................4
\item Chern-Dimer mapping for anisotropic lattices.....................................6
\item Chern numbers in time of flight measurements...................................8
\item Incommensurate flux: Chern tree.....................................................11
\item Quantum spin-Hall effects with ultracold atoms..............................15
\item Open questions and outlooks..........................................................18
\end{itemize}

\section*{Ultracold Atoms, synthetic gauge fields, and topological states}

Quantum engineering, i.e., preparation, control and detection of quantum systems,
at micro- to nano-Kelvin temperatures has turned ultracold atoms into a versatile tool for
discovering new phenomena and exploring new horizons in diverse branches of physics.
Following the seminal experimental observation of Bose Einstein condensation,
simulating many-body quantum phenomena using cold atoms, which in a limited sense 
realizes Feynman's dream of quantum simulation, has emerged as an active frontier at the 
cross section of AMO and condensed matter physics.
There is hope and excitement that these highly tailored and well controlled systems
may solve some of the mysteries regarding strongly correlated electrons, 
and pave ways for the development of quantum computers.

The subject of this chapter is motivated by recent breakthroughs in creating ``artificial gauge fields''
that couple to neural atoms in an analogous way that electromagnetic
fields couple to charged particles \cite{Ian-mag}. This opens up the possibility of 
studying quantum Hall (QH) states and
its close cousin quantum spin Hall (QSH) states, 
which belong to the family of exotic states of matter
known as {\it Topological Insulators}.

QH and QSH effect usually requires magnetic field and spin-orbit coupling, respectively.
In contrast to electrons in solids, cold atoms are charge neutral and interact with
external electromagnetic field primarily via atom-light coupling. Artificial orbital magnetic
field can be achieved by rotating the gas \cite{cooper}. In the rotating frame,
the dynamics of atoms becomes equivalent to charged particles in a magnetic field
which is proportional to the rotation frequency. To enter the quantum Hall regime using
this approach requires exceedingly fast rotation, a task up to now remains a technical
challenge. In this regard, another approach to engineer artificial magnetic field
offers advantages. This involves engineering the (Berry's) geometric phase of atoms
by properly arranging inhomogeneous laser fields that modify the eigenstates (the
dressed states) via atom-light coupling \cite{id}. Such geometric phase can mimic for example the
Aharonov-Bohm phase of orbital magnetic field. Because the dressed states can be
manipulated on scales of the wavelength of light, the artificial gauge fields can be
made very strong. For example, the phase acquired by going around the lattice unit cell
can approach the order of $\pi$. This gives hope to enter the quantum Hall regime.
The strength of this approach also lies in its versatility. Artificial gauge fields in a multitude of forms,
either Abelian (including
both electric and magnetic field) or non-Abelian (including effective spin-orbit coupling),
can be introduced for continuum gases or atoms in optical lattices \cite{id}.
Recent NIST experiments have successfully demonstrated artificial magnetic field \cite{Ian-mag} as well as
spin-orbit coupling \cite{Ian-so} for bosonic Rb atoms and experiments are underway to reach QH and QSH regimes.

\begin{figure}
{\includegraphics[width=\textwidth]{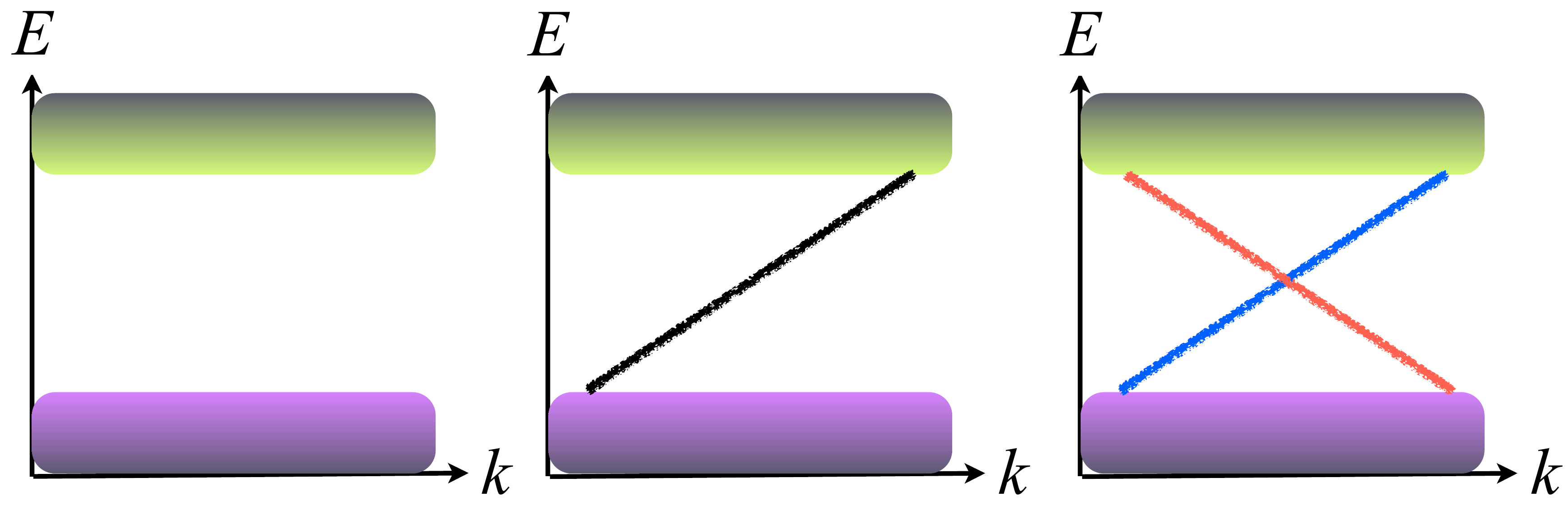}}
\leavevmode \caption{ Figure illustrates distinction between ordinary band insulator (left), QH (middle) 
and QSH (right) insulators in the energy spectrum. Connecting the conduction and 
valence bands are the edge states.}
\label{edge}
\end{figure}

\section*{Topological aspects: Chern numbers and edged states }

Topological insulators (TI) are unconventional states of matter \cite{rmp} 
that are insulating in the bulk but conduct along the boundaries. A more 
general working definition of TI, applicable to the case of neutral atoms, is 
that they are band insulators with a bulk gap 
but robust gapless boundary (edge or surface) modes ( Fig. ~\ref{edge} ). 
These boundary modes are protected against
small deformations in the system parameters, as long as generic symmetries 
such as time-reversal or charge-conjugation are preserved and the
bulk gap is not closed. Such states arise even in systems of non-interacting fermions and
are considerably simpler than topological states of strongly interacting
electrons, such as the fractional QH effect. Despite such deceiving simplicity, 
as we will show in this chapter, the subject 
is very rich and rather intricate, with plenty of open questions.

We will consider two prime examples of TIs in two dimension,
the QH and QSH states. 
A topological insulator is not characterized
by any local order parameter, but by a topological number
that captures the global structures of the wave functions.
For example, associated with each integer quantum Hall state is an integer number,
the Chern number . 
The Chern number also coincides with the quantized Hall conductance (in
unit of $e^2/\hbar$) of the integer QH state.
The quantization of the Hall conductance is extremely robust,
and has been measured to one part in $10^9$ \cite{Klitzing}.
Such precision is a manifestation of the topological nature of
quantum Hall states.
A quantum Hall state is accompanied by chiral edge modes, that propagate along the edge along one direction
and their number
equals the Chern number.
Semi-classically, these edge states can be visualized as resulting from the
skipping orbits of electrons bouncing off the edge while 
undergoing cyclotron motion in external magnetic field.
The edge states are insensitive to disorder because there are no states available 
for backscattering, a fact that underlies the perfectly quantized electronic 
transport in the QHE.
This well known example demonstrates the ``bulk-boundary
correspondence'', a common theme for all topological insulators. 

The QSH state is time-reversal invariant \cite{qsh}. It is characterized by a binary $Z_2$ 
topological invariant $\nu=1$. 
The edge modes in this case come in the form of Kramers pairs, with one set of modes being the time-reversal
of the other set. In the simplest example, spin up electron and spin down electron travel
in opposite direction along the edge. A topologically nontrivial state with $\nu=1$
has odd number of Kramers pairs at its edge. Topologically trivial states have $\nu=0$,
and even number of Kramers pairs, the edge states in this case are not topological stable. 
This is another manifestation of the ``bulk-boundary correspondence".
Recent theoretical research has achieved a fairly complete
classification, or ``periodic table'', for all possible topological band insulators and topological
superconductors/superfluids in any spatial dimensions \cite{ten,kita,qi_field}. 
Several authoritative reviews of this rapidly evolving field now exist 
\cite{today,rmp,Qi-zhang-rev}, which also cover the fascinating subject
of how to describe TIs using low energy effective field theory with topological
(e.g., Chern-Simons or $\theta$-) terms.
\\
 
In cold atom systems, transport measurement
such as Hall conductance is difficult. Also, the existence of an overall trap potential instead of a
sharp edge for a solid state sample, as well as the lack of a precision local probe such as STM,
makes it hard to probe the edge states. Therefore, identifying topological state such as
a QH state and extract the Chern number remains a challenging task. Here we show that
Chern numbers are ``hiding'' in time of flight images of the cold atom systems, a routine measurement in cold atom
experiments.
\\

\section*{The Hofstadter model and the Chern number}

\begin{figure}
{\includegraphics[width=1.0\textwidth]{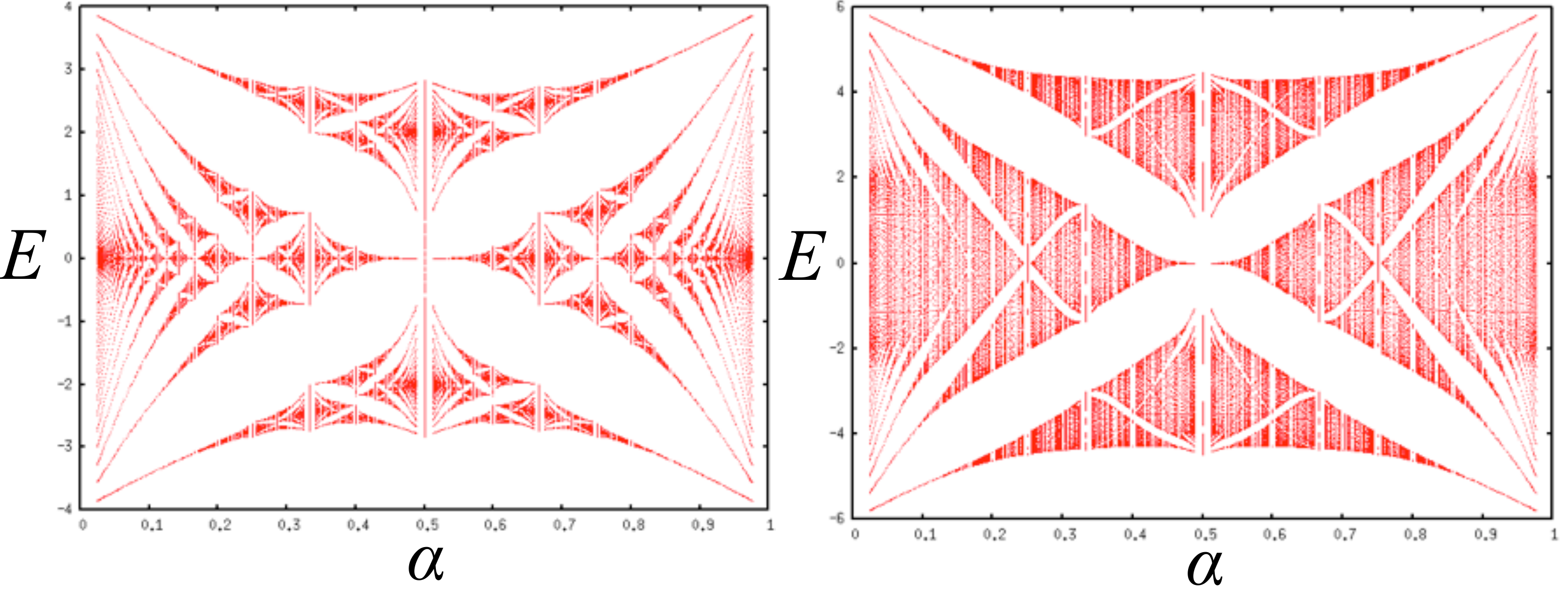}}
\leavevmode \caption{ The energy spectrum as function of the magnetic flux $\alpha$ per plaquette, commonly referred as the butterfly spectrum, for the Hofstadter model with $\lambda=t_y/t_x=1,2$ respectively.}
\label{butt}
\end{figure}

For a moderately deep square optical lattice, the dynamics of (spinless) fermionic atoms
in the presence of an artificial magnetic field is described by the tight binding Hamiltonian
known as the Hofstadter model \cite{Hof},
\begin{equation}
H=-\sum_{nm} t_x c_{n+1,m}^{\dagger} e^{i\theta_{n+1,m:n,m}} c_{n,m}+
t_y c_{n,m+1}^{\dagger} e^{i\theta_{n,m+1:n,m}} c_{n,m}+h.c.
\label{Hamil}
\end{equation}
where $c_{n,m}$ is the fermion annihilation operator at site $(n,m)$, with $n/m$ 
is the site index along the $x/y$ direction,
$t_x$ and $t_y$ are the nearest-neighbor hopping
along the $x$ and the $y$-direction respectively. The anisotropy in hopping is described
by dimensionless parameter $\lambda=t_y/t_x$.
The phase factor $\theta_{n,m:n',m'}$ is given the line integral
of the vector potential ${\bf A}$ along the link from site $(n',m')$ to
$(n,m)$, $(2\pi e/ch) \int {\bf A}\cdot d{\bf l}$. 
The magnetic flux per plaquete measured in flux quantum is labeled
$\alpha$, and represents the strength of magnetic field perpendicular to 
the 2D plane. The energy spectra for $\lambda=1,2$ are shown
in Fig. \ref{butt}. At commensurate flux, i.e., $\alpha$ is a rational number $\alpha=p/q$
with $p$ and $q$ being integers, the original energy band 
splits into $q$ sub-bands by the presence of magnetic field, and the spectrum has $q-1$ gaps.
For incommensurate flux, namely when $\alpha$ is an irrational number, the fractal spectrum
contains a hierarchy of sub-bands with infinite fine structures. As shown in  Fig. \ref{butt},
varying anisotropy changes the size of the gaps without closing any gap.

In the Landau gauge, the vector potential
$A_x=0$ and $A_y = -\alpha x$ and
the unit cell is of size $q\time 1$ and the magnetic Brillouin zone
is $ -\pi/q \le k_x \le \pi/q$ and $-\pi \le k_y \le \pi$ (the lattice spacing is set to be 1).
The eigenstates of the system
can be written as $\Psi_{n,m}= e^{ik_xn+ik_y m} \psi_n $ where $\psi_n$ satisfies the Harper
equation \cite{Hof},
\begin{equation}
e^{ik_x}\psi^r_{n+1}+e^{-ik_x} \psi^r_{n-1} + 2\lambda \cos ( 2 \pi n \alpha+ k_y)\psi^r_n = -E \psi^r_n .
\label{harper}
\end{equation}
Here $\psi^r_{n+q} =\psi^r_n$, $r=1, 2, ...q$, labels linearly independent solutions.
The properties of the two-dimensional system can be
studied by studying one-dimensional equation (\ref{harper}).
For chemical potentials inside each energy gap, the system
is in an integer quantum Hall state characterized by its Chern number
$c_r$ and the transverse conductivity $\sigma_{xy}=c_r {e^2}/{h}$, where the Chern number of $r$-filled bands is given by,
\begin{equation}
c_r= \mathrm{Im}\sum_{r=1}^{l} \int dk_x dk_y\sum_{n=1}^q\partial_{k_x}(\psi^r_n)^* \partial_{k_y} \psi^r_n.
\label{kubo}
\end{equation}
Here the integration over $k_x,k_y$ is over the magnetic Brillouin zone. 

\begin{figure}
{\includegraphics[width=1.0\textwidth]{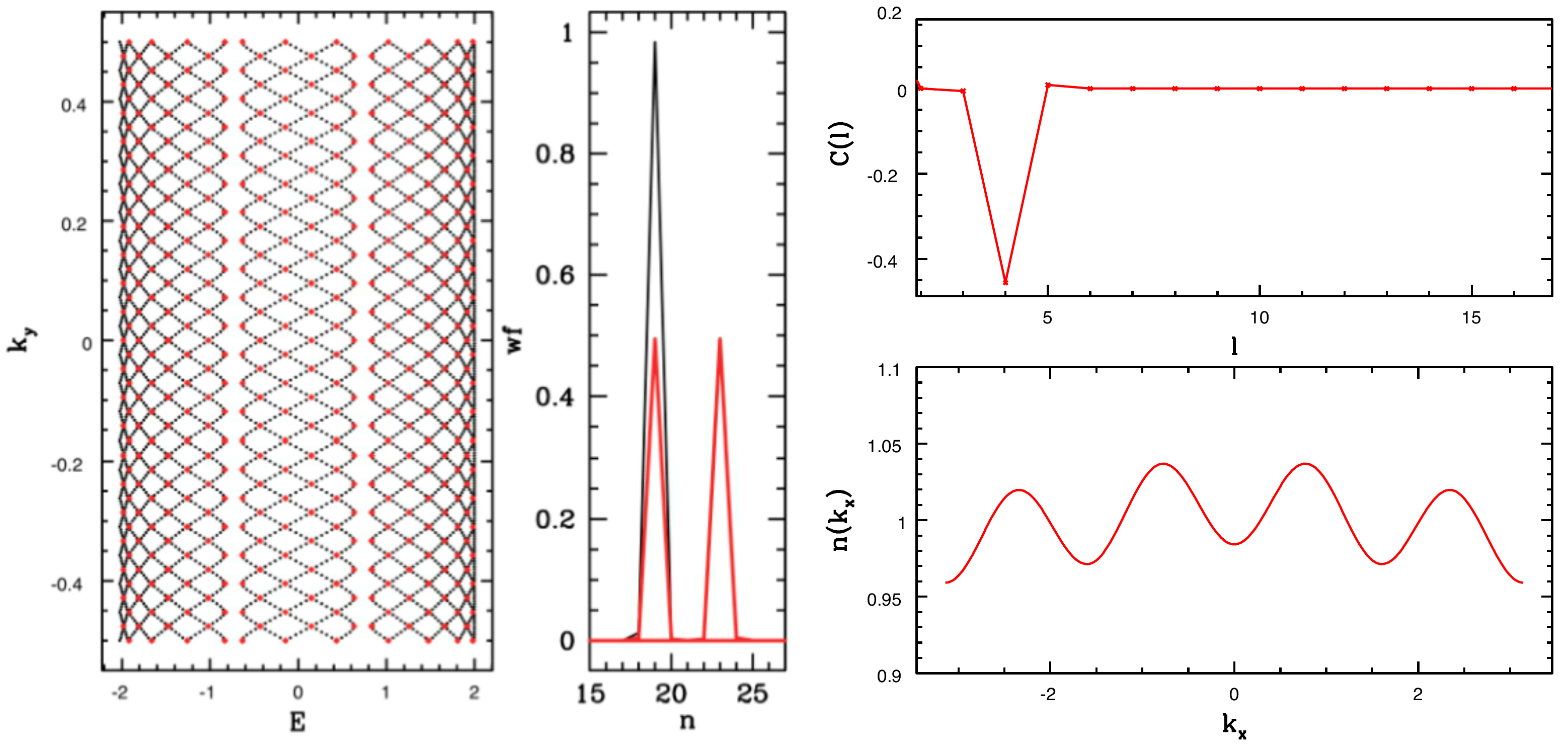}}
\leavevmode \caption{Left: energy spectrum for $\lambda=10$, $\alpha=13/21$ . The red dots
indicate dimerized states at the band edges. Middle: the wave function of a state localized at a single site,
with energy away from the band edge (black), and the dimerized wave function that localizes at two distinct sites
at a distance of four-lattice spacing apart for state at band edge near a gap with Chern number $4$ (red).
Figure on the right shows the two point correlation function $C(l)$ that vanishes unless
$l=4$ resulting in almost sinusoidal momentum distribution with period $4$.}
\label{dimer}
\end{figure}

\section*{Chern-Dimer mapping for anisotropic lattices }
As illustrated in Fig. (\ref{butt}), the hopping anisotropy changes the size of the gaps,
without ever closing the gap. Hence the topological aspects of the underlying states
are preserved, as $\lambda$ is continuously tuned. In the limit 
$\lambda \gg 1$, the problems becomes tractable using degenerate perturbation theory~\cite{fradkin91}.
For $\lambda\rightarrow\infty$, the $q$ bands have dispersion $E_r(k_y)= -2 t_y \cos( 2\pi \alpha r+k_y)$,
$r\in\left\{1,\ldots, q\right\}$. These eigenstates are extended along 
$y$ but localized to single sites as a function of $x$, with wave function
$\psi^r_{n}=\delta_{n,r}$. These cosine bands shifted from each other intersect at 
numerous degeneracy points ${k^*_y}$. 
For finite $\lambda$, the hopping term proportional to $t_x$ hybridizes the two states at each crossing, 
resulting an avoided crossing  as shown in Fig~ (\ref{dimer}).
Near each degeneracy point the system is well described by 
a two-level effective Hamiltonian~\cite{fradkin91}, 
\begin{equation}
{H}(\mathbf{k})=\Delta_{\ell}[\cos(k_x\ell){\sigma}_x+\sin(k_x\ell){\sigma}_y]
+v_{\ell}(k_y-k^*_y){\sigma}_z.
\label{two-level}
\end{equation}
Here ${\sigma}_{x,y,z}$ are the Pauli matrices in the 
space of localized states $\ket{r}$ and $\ket{r^\prime}$, and $\ell = r^\prime-r$
is the spatial separation between these two localized states. The expression for 
$\Delta_{\ell}$ and $v_{\ell}$ are given in Ref. \cite{chernPRA}. 
The eigenstates are equal-weight superpositions of $\ket{r}$ and $\ket{r^\prime}$,
\begin{equation}
\psi^\pm_{n}=\frac{1}{\sqrt{2}}\left(\delta_{n,r}+e^{i\beta_\pm}\delta_{n,r^\prime}\right),
\label{dimerwf}
\end{equation}
where $\beta_{+}=-\ell(k_x +\pi)+\pi$ and $\beta_{-} = -\ell(k_x +\pi)$ are relative phases for the upper and lower band edges, respectively. We call such dimerized state ``Chern-dimers''. 
Starting from Eq.~(\ref{dimerwf}), it is straightforward to show~\cite{fradkin91} that 
the corresponding Chern number is related to the phase factor $e^{i k_x\ell}$, and simply given by
$c_r=\ell$. Thus, for sufficiently anisotropic hopping, dimerized states form at the
band edges with its spatial extent along $x$ equal to the Chern number of the 
corresponding gap. Such one-to-one correspondence between the Chern number and 
the size of the real space dimer (in the Landau gauge) will be referred to as ``Chern-dimer mapping''.
While calculation of Chern number in this limit is not new, this mapping
has not been elucidated and emphasized until our work \cite{chernPRA}. 

The formation of Chern-dimers of size $\ell=c_r$ at the edges of the $r$-th gap implies 
a ``hidden spatial correlation''. The correlation function 
\begin{equation}
C(l) =\sum_{n,m}\langle c^\dagger_{n,m} c_{n+l,m}\rangle 
\end{equation}
will peak at $l=\ell=c_r$. For chemical potential in the $r$-th gap, as $\lambda\rightarrow \infty$,
( Fig. ~\ref{dimer})
$C(l)$ asymptotes to a delta function $C(l)\rightarrow \delta(l-\ell)$, due to
the contribution of Chern-dimers at the lower edge of the $r$-th gap. Note that 
the net contribution to $C(l)$ from all other 
dimerized states associated with gaps fully below the Fermi energy is zero,
because $\beta_+-\beta_- = \pi$ for each of these crossings. 
Remarkably, $C(l)$, or more precisely its Fourier transform $n(k_x)$,
is directly accessible in the time-of-flight (TOF) image, an routine observable
in cold atom experiments, 

In the time-of-flight experiments, the artificial gauge field is first turned off
abruptly. This introduces an impulse of effective electric field that converts
the canonical momentum of atoms (when gauge field was on) into the mechanical momentum
(these subtleties are discussed in \cite{chernPRA}).
Then the optical lattice and trap potential are turned off, the
density of the atomic cloud $n(\mathbf{x})$ is imaged after a period of free expansion $t_{\rm TOF}$. 
The image
is a direct measurement of the (mechanical) momentum distribution function with ${\bf k}= m{\bf  x}/\hbar t_{\rm TOF}$,
\begin{equation}
n(\mathbf{k})=\sum_{\mathbf{r,r'}}\langle c^\dagger_{\mathbf{r}} c_{\mathbf{r'}}\rangle e^{i\mathbf{k}\cdot(\mathbf{r-r'})},\label{MomentumDistribution}
\end{equation}
where the site label $\mathbf{r}=m\hat{x}+n\hat{y}$, and 
for simplicity we neglect the overall prefactor and the Wannier function envelope. Thus, 
from TOF images, the 1D momentum distribution can be constructed by integrating $n({\bf k})$ over $k_y$, 
$n(k_x)=\int_{-\pi}^{\pi} d k_y {n}({\bf k})$,
which is nothing but the Fourier transform of $C(l)$. Thus, in the limit of large $\lambda$,
$C(l)$ becomes a delta function, and the 1D momentum distribution function becomes 
\begin{equation}
n(k_x)= r\pm\frac{1}{2}\cos(c_r k_x) +...
\label{nkx-asymp}
\end{equation}
It assumes a sinusoidal form, with oscillation period given by the Chern number. Note the 
sign depends on which gap the chemical potential lies in \cite{chernPRA},
and the off-set $r$ comes from the contribution from the filled bands below the Fermi energy.
In summary, we have arrived at the conclusion that Chern numbers can be simply read off
from $n(k_x)$ as measured in TOF, in the large anisotropy limit.

\section*{Chern number in time-of-flight images}

We now address the question whether it is still possible to extract $c_r$ from TOF when away from 
the large $\lambda$ limit. Note that, while the Chern number does not change inside the same gap 
as $\lambda$ is tuned, $n(k_x)$ is not topologically invariant. There is a prior no 
reason to expect Chern number continues to leave its fingerprints in $n(k_x)$. Motivated by these considerations,
we have carried out detailed numerical studies of momentum distribution in the entire
$\alpha-\lambda$ parameter plane, going
well beyond the ``Chern-dimer regime''. Fig. ~(\ref{dimer}) highlights our results
with various $\lambda$ but fixed $\alpha$ and Chern number.
As $\lambda$ is reduced, $n(k_x)$ gradually deviates from the sinusoidal distribution. Yet,
the local maxima (peaks) persist. In the limit of isotropic lattice, $n(k_x)$ shows
pronounced local peaks. For a wide region of parameters, we can summarize our finding
into an empirical rule: 
{\it the number of local peaks in $n(k_x)$ within the first Brillouin zone is equal to the Chern number}.
Note that this rule is not always precise, as seen in the Fig. (\ref{dimer}) for the case of Chern number
$3$ for intermediate values of $\lambda$ where local peaks become a broad shoulder-like structure. 
Despite this, it is valid in extended regions including
both large $\lambda$ and $\lambda=1$. In fact, given the reasoning above,
our empirical rule seems to work surprisingly well.
But this is an encouraging result for cold atoms experiments. From these
distinctive features in $n(k_x)$, the Chern number can be determined rather
easily by counting the peaks, especially for the major gaps with low Chern number.

\begin{figure}
{\includegraphics[width=1.0\textwidth]{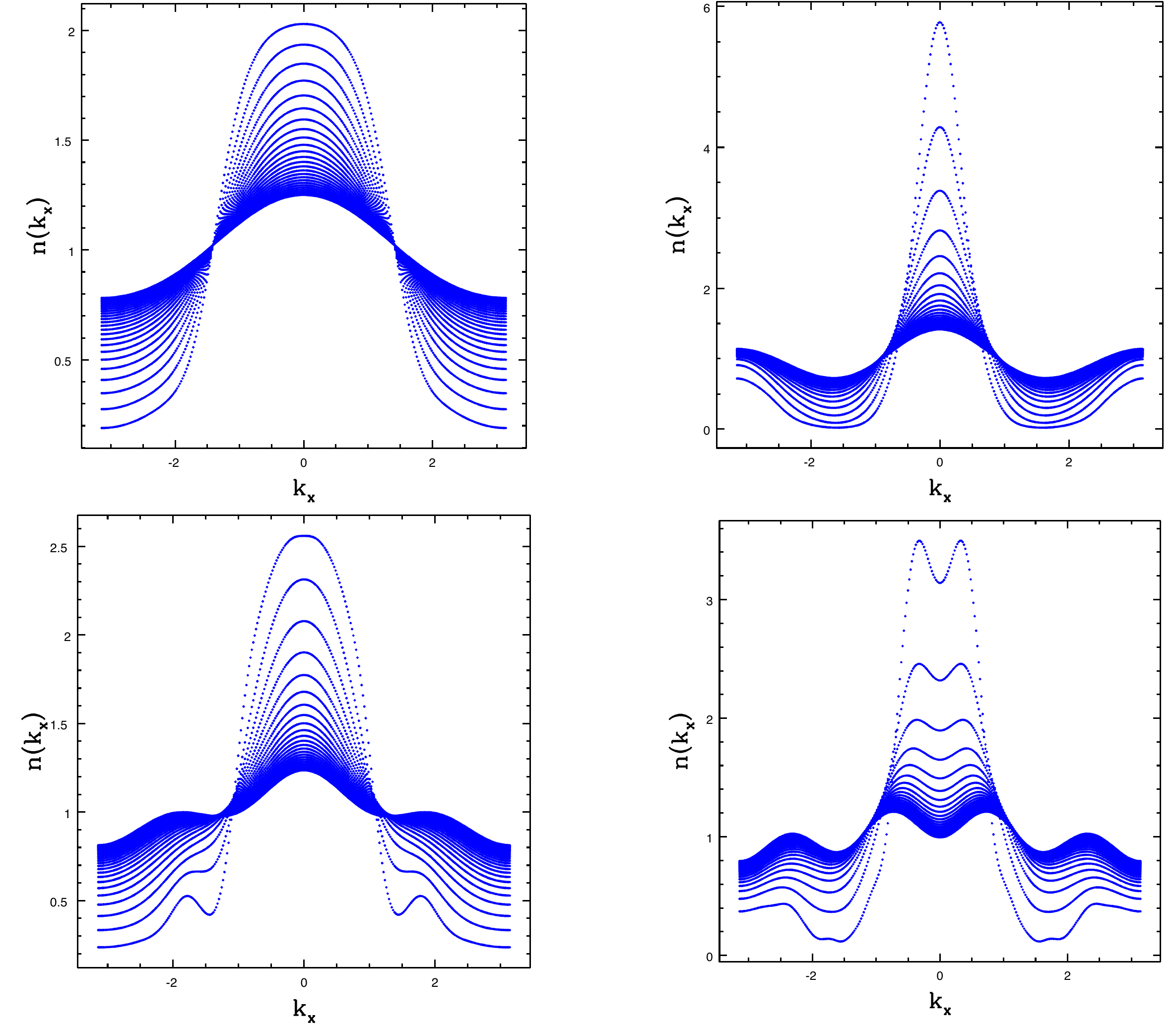}}
\leavevmode \caption{ Momentum distribution as anisotropy parameter varies from $1-10$,
for $\alpha=.45$ for chern numbers $1$ (top-left), $2$ (top-right), $3$ (bottom-left), and $4$ (bottom-right).}
\label{dimer}
\end{figure}

\begin{figure}
{\includegraphics[width=\textwidth]{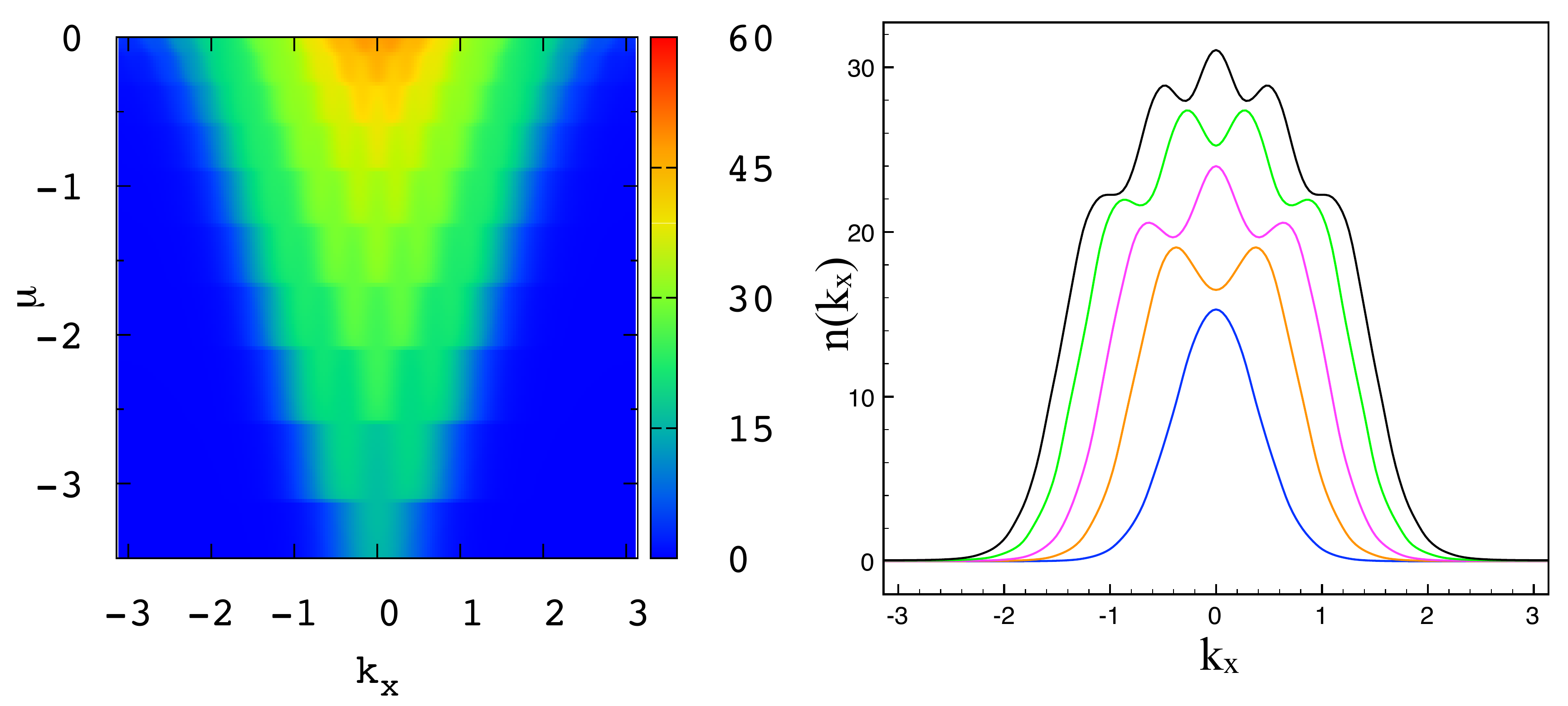}}
\leavevmode \caption{ Left: $n(k_x,\mu)$ in false color for a $50\times 50$ lattice
with isotropic hopping and flux $\alpha=0.05$. Right: $n(k_x)$
for chemical potential $\mu/t$=-3.4, -2.8, -2.4, -1.8, -1.5 respectively (bottom to top).
The corresponding Chern number is 1, 2, 3, 4, and 5.}
\label{LandauL}
\end{figure}

The Chern-dimer mapping provides a simple, intuitive understanding of $n(k_x)$ 
in the limit of large $\lambda$. For the isotropic case, there is no dimerization, 
why are there pronounced peaks in $n(k_x)$, and why is the peak number precisely equal to
the Chern number? To obtain further insight, we considered the low flux limit, $\alpha\rightarrow 0$,
where the size of the cyclotron orbit is large compared to the lattice spacing,
and the system approaches the continuum limit.
The spectrum then can be approximated by discrete Landau levels.
For $r$ filled Landau levels, $n(k_x)$ can be obtained rather easily,
\begin{equation}
n(k_x)=\sum_{\nu=0}^{r-1} (2^\nu \nu !)^{-1}H^2_\nu({k_x})e^{-k^2_x}, \label{landau}
\end{equation}
where $H_{\nu}$ is the Hermite polynomial, and $k_x$ is measured in the inverse magnetic length $\sqrt{eB/\hbar c}$.
$n(k_x)$ is given by the weighted sum of consecutive positive-definite function, $H^2_\nu({k_x})$.
The node structure of $H_{\nu}$ is such that $n(k_x)$ has {\it exactly} $r$ local maxima (peaks)
for $r$ filled Landau levels. This nicely explains the numerical results for 
the Hofstadter model at low (but finite) flux $\alpha=1/q$, where
the Chern number $c_r=r$ in the $r$-th gap, and $n(k_x)$ has precisely $r$ peaks (see Fig. \ref{LandauL}).
Note that this calculation does not directly explain the feature shown in Fig. \ref{dimer} for
$\alpha=0.45$ and $\lambda=1$. As $\alpha$ increases, for $\mu$ residing within the same energy gap, 
the distinctive features of $n(k_x)$ are generally retained, so the counting rule is still valid. 
Compared to the low flux regime, however,
the local peaks (except the one at $k_x=0$) move to higher momenta.

\section*{Incommensurate flux: Chern tree}

While it is numerically straightforward to investigate the $n(k_x)$ or
$\sigma_{xy}$ for QHE at incommensurate flux $\alpha$, analytical analysis of the Chern number 
for this case represents a conceptual as well as a technical challenge. In fact, for 
irrational $\alpha$, the system no longer has translational symmetry,
and the notion of magnetic Brillouin zone or Bloch wave function
lose their meaning. The incommensurability 
acts like {\it correlated disorder}. The TKNN formula for a periodic system
is not directly applicable. In addition, the energy spectrum becomes 
infinitely fragmented (a Cantor set). One expects then a dense distribution of 
Chern numbers even within a small window of energy. Then two questions arise.
First, will these Chern numbers largely random or are there deterministic 
patterns governing their sequence as $\mu$ is continuously varied? Second,
what is the experimental outcome of $n(k_x)$ or $\sigma_{xy}$ when measured
with a finite energy resolution?

\begin{figure}
{\includegraphics[width=1.0\textwidth]{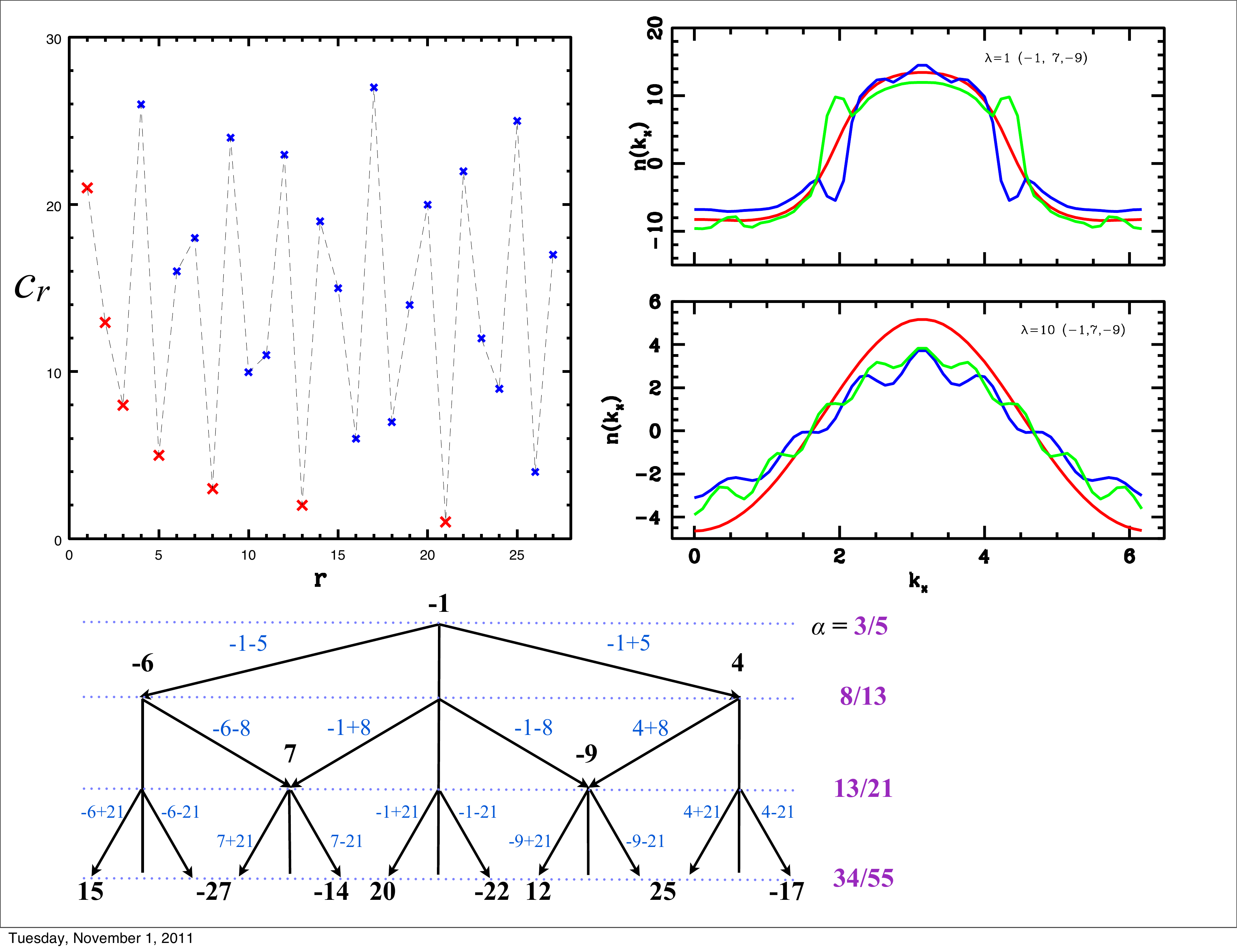}}
\leavevmode \caption{The Chern tree at golden-mean flux showing
four consecutive rational approximations:
$\alpha=3/5, 8/13,13/21, 34/55$ starting with Chern number $-1$. 
The horizontal axis is not to scale and only shows the correct order of Chern numbers. Top-left figure
shows Fibonacci (red) and non-Fibonacci (blue) Chern numbers vs the gap index $r$
for $\alpha=34/55$. Top-right shows momentum distribution 
for three different topological states characterized
by Chern numbers $1$ (red), $7$ (blue) and $-9$ (green). Figure illustrates self-averaging aspect of the
momentum distribution:
states $7$ and $-9$ that reside in close proximity to the state with Chern number unity 
have ripples that tend to cancel each other.}
\label{Chtree}
\end{figure}

We address some of these questions here for the isotropic lattice ($\lambda=1$)
using the golden-mean flux, $\alpha=(\sqrt{5}-1)/2$, as an example. We will
employ a well known trick, and approach the irrational limit 
by a series of rational approximation $\alpha_n=F_n/F_{n+1}$,
where $F_n$ is the $n$-th Fibonacci numbers. Explicitly, the series
$3/5, 5/8, 8/13, 13/21, 21/34, 34/55, 55/89...$ approaches the golden-mean  
with increasing accuracy. At each stage of the rational
approximation, the energy spectrum and the corresponding sequence of 
Chern numbers can be determined.  
Higher Chern numbers are typically associated to smaller gaps in the asymptotically
fractal spectrum. 
At first sight, the Chern number associated with the $r$-th gap seems an 
irregular function of $r$ as shown in Fig.~(\ref{Chtree}) for $\alpha=34/55$.
This is in sharp contrast with the case of $\alpha=1/q$, where $c_r = r$ 
for the $r$-th gap.
 
Now we show analytically that there actually exists a deterministic pattern underlying
the sequence of Chern numbers. This pattern follows from the solutions of the Diophantine
 equation (DE)\cite{TKKN},
$r=c_r p + s_r q$, where $r$ is the gap index,
$c_r$ is the Chern number, at each level of the rational approximation, $p=F_n,q=F_{n+1}$.
We first classify all the Chern numbers into two classes, the
Fibonacci Chern numbers and the non-Fibonacci Chern numbers.
All the solutions to the DE, which determine the Chern number, 
can be exhausted by following two simple rules, labeled as
I, II and explicitly stated below.

(I)  Fibonacci Chern numbers occur at gaps labeled by
a Fibonacci index $r=F_m$: for $\alpha=F_{n-1}/F_n$, $c_{r=F_m}= -(-1)^{n-m} F_{n-m}$ .
Here $r=1$ corresponds to the first gap near the band edge with Chern number $F_{n-2}=(-1)^{n_f}(q-p)$, while
the last Fibonacci gap near the band center has a Chern number $(-1)^{n_f}$
where $n_f$ counts the number of Fibonacci numbers less than $|{q}/{2}|$.
The Fibonacci Chern numbers increases
monotonically in magnitude as one moves from the center towards the edge of the 
band (see Fig.~\ref{Chtree}).

(II) The rest are the non-Fibonacci Chern numbers.
The Chern number for the $r$-th gap
and its neighboring gaps ($r\pm1$) obey the recursion relation,
\begin{eqnarray*}
c_r &=& {(c_{r+1}+c_{r-1})}/{2},\\
c_{r+1}&=&c_r+(-1)^{n_f}(q-p),\\
c_{r-1}&=&c_r-(-1)^{n_f}(q-p),
\label{RR}
\end{eqnarray*}

Following these rules, we can construct an exhaustive list of all Chern numbers
at each level of rational approximation. 
We have numerically checked that the Chern numbers computed from the TKNN formula
are consistent with the predictions from the DE above.
This procedure yields a hierarchy
of Chern numbers, which can be neatly organized into a tree structure, thanks to
the recursion relation above.
A subset of this tree hierarchy is shown in Fig. \ref{Chtree}.
Most notably, we find the average of the left and right neighboring
Chern number is exactly the Chern number in the middle.

This property helps to understand the measurement outcomes of real experiments,
which inevitably involves average of states with energy very close to each other.
In the context of cold atoms, due to the overall trap potential, the local chemical
potential slowly varies in space. These local contributions sum up in TOF images
according to the local density approximation. The self-averaging property 
to some degree ensures that such average yields definite result, and the
infinitely fine details in the energy spectrum 
are largely irrelevant. This aspect is illustrated in Fig. \ref{Chtree} for the
Chern tree shown in Fig. \ref{Chtree}.
The momentum distribution $n(k_x)$ in the average sense for 
$\alpha=3/5, 8/13, 13/21, 34/55$ will 
be essentially the same.

\begin{figure}
{\includegraphics[width=\textwidth]{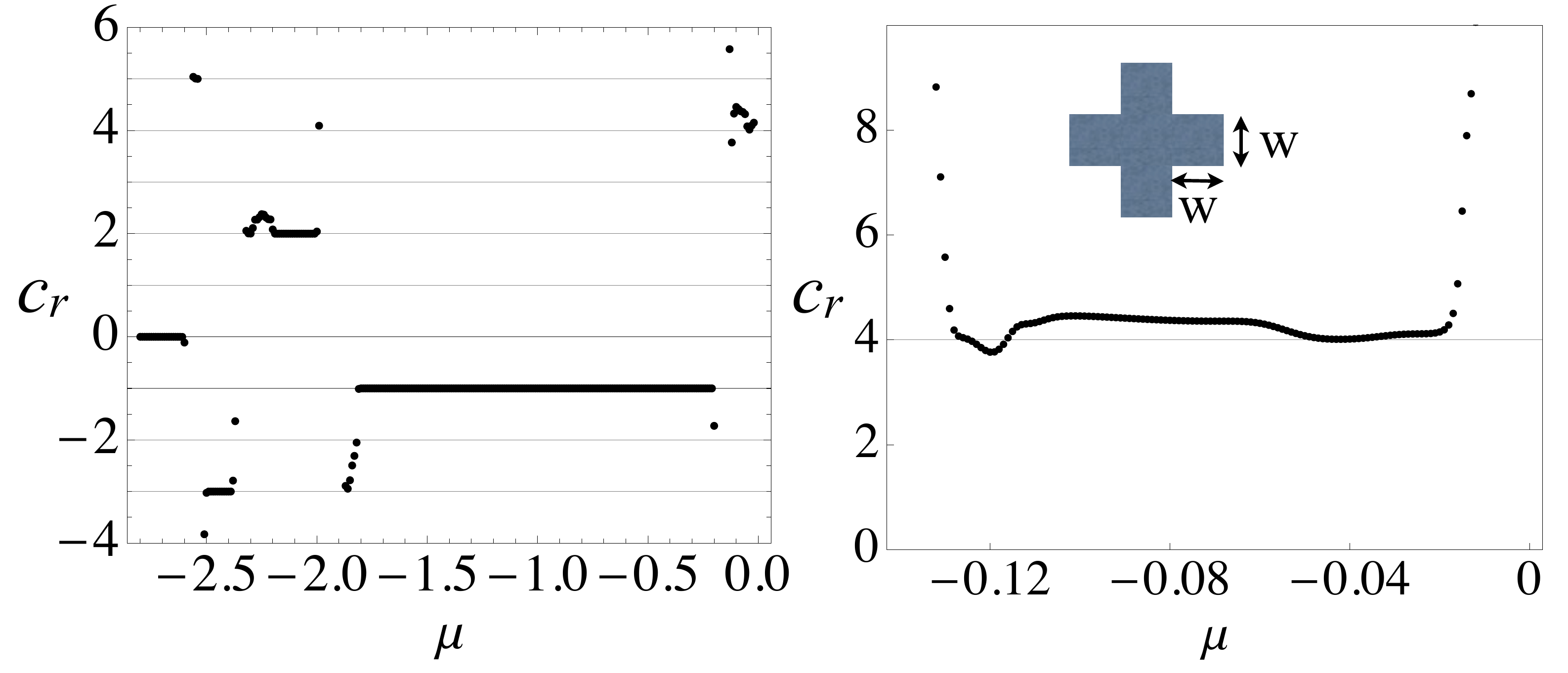}}
\leavevmode \caption{Left: the quantized conductance $c_r=\sigma_{xy}/(e^2/h)$ versus the chemical potential $\mu$
for $\lambda=1$ and $\alpha=(\sqrt{5}-1)/2$. Right: the blow up for $\mu$ close to zero showing
an extended plateau at $c_r=4$. $c_r$ is computed for a four-terminal quantum Hall bar
of width $W=20a$ (inset). The device is connected to four semi-infinite leads of the same material with
inelastic scattering rate $0.005t$. }
\label{4term}
\end{figure}

Similar consideration applies to $\sigma_{xy}$. One may see a robust plateau, which may not be
perfectly quantized, in a region of chemical potential due to the self-averaging.
We have numerically computed $\sigma_{xy}$ using the
Landauer-Buttiker formalism \cite{Butti} for
a mesoscopic quantum Hall bar (inset of Fig. \ref{4term}) at golden-mean flux and connected
to four semi-infinite leads. The transmission coefficients $T_{ij}$ are
obtained by solving for the Green's function of the sample
and the level width functions of each leads.
The transverse conductivity at golden-mean flux is shown in Fig. \ref{4term}.
One clearly sees a conductance plateau with $c_r=4$, in addition to $c_r=1,2,3$.

\section*{Quantum spin Hall effect for ultracold atoms}

In solid state, the QSH effect is closely related to spin-orbit coupling.
In a recent study \cite{goldman}, it was shown that the effect of
spin-orbit coupling can be achieved for cold atoms by creating $SU(2)$ 
artificial gauge fields.

\begin{figure}
\includegraphics[width=\textwidth]{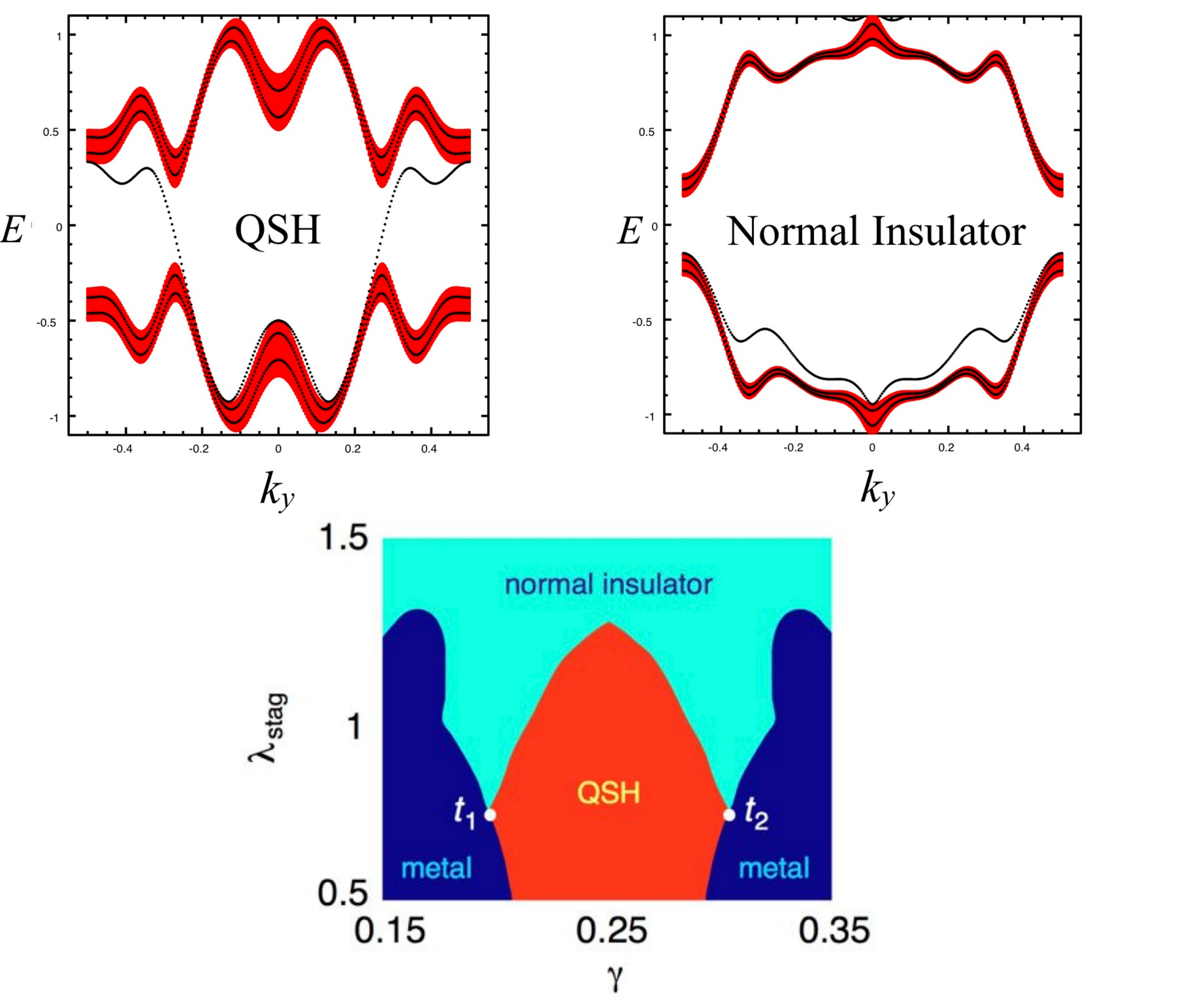}
\leavevmode \caption { Shows the bulk (red) and surface spectrum (black) , near half-filling,
for $\alpha=1/6$, $\gamma=.25$, $\lambda=1$ , and $\lambda_{stag}=.5$ ( left) and $\lambda_{stag}=1.5$ ( right). The presence of gapless
edge modes for $\lambda_{stag}=.5$ and their absence for $\lambda_{stag}=1.5$ shows that these two respectively correspond to topologically
non-trivial and trivial phases. The bottom caption shows
the corresponding Phase diagram in $\gamma-\lambda_{stag}$
plane .}
\label{qshiso}
\end{figure}

The theoretical model that captures the essential properties of this system is given by a Hamiltonian describing
fermions hopping on a square lattice in the presence of a SU(2) gauge field
\begin{eqnarray}
\mathcal{H}=-t\sum_{mn}  \bf{c}_{m+1,n}^{\dagger} e^{i \theta_{x}}  \bf{c}_{m,n}
+ \bf{c}_{m,n+1}^{\dagger} e^{i \theta_y}  \bf{c}_{m,n}+ h.c.+
\lambda_{stag} (-1)^{m} \,  \bf{c}^{\dagger}_{m,n}  \bf{c}_{m,n}
\label {2dh}
\end{eqnarray}
where $ \bf{c}_{m,n}$ is the 2-component field operator defined on a lattice site $(m,n)$,
and $\lambda_{stag}$ is the onsite staggered potential along the $x$-axis.
The phase factors $\theta_{m,n}$ corresponding to $SU(2)$ gauge fields are chosen to be,
\begin{eqnarray}
\theta_x &=& 2\pi \gamma {\bf \sigma_x}, \nonumber \\
\theta_y &=&  2\pi x \alpha {\bf \sigma_z},
\label{gaugefields}
\end{eqnarray}
where ${\bf \sigma_{x,z}}$ are Pauli matrices.
The control parameter $\gamma$ is a crossover parameter between QH and QSH states as $\gamma=0$ 
describes uncoupled double QH systems while $\gamma=\pi/2$ corresponds to the two maximally coupled QH states.
It should be noted that the Hamiltonian (\ref{2dh}) is invariant under time-reversal.
The two components of the field operators correspond in general to a pseudo-spin 1/2,
but could also refer to the  spin components of atoms, such as $^6$Li,  in a $F=1/2$ ground state hyperfine manifold.
For $\gamma=0$ case, spin is a good quantum number and
the system decouples into two subsystems which are respectively subjected to the magnetic fluxes $\alpha$ and $-\alpha$.

\begin{figure}
\includegraphics[width=1.0\textwidth]{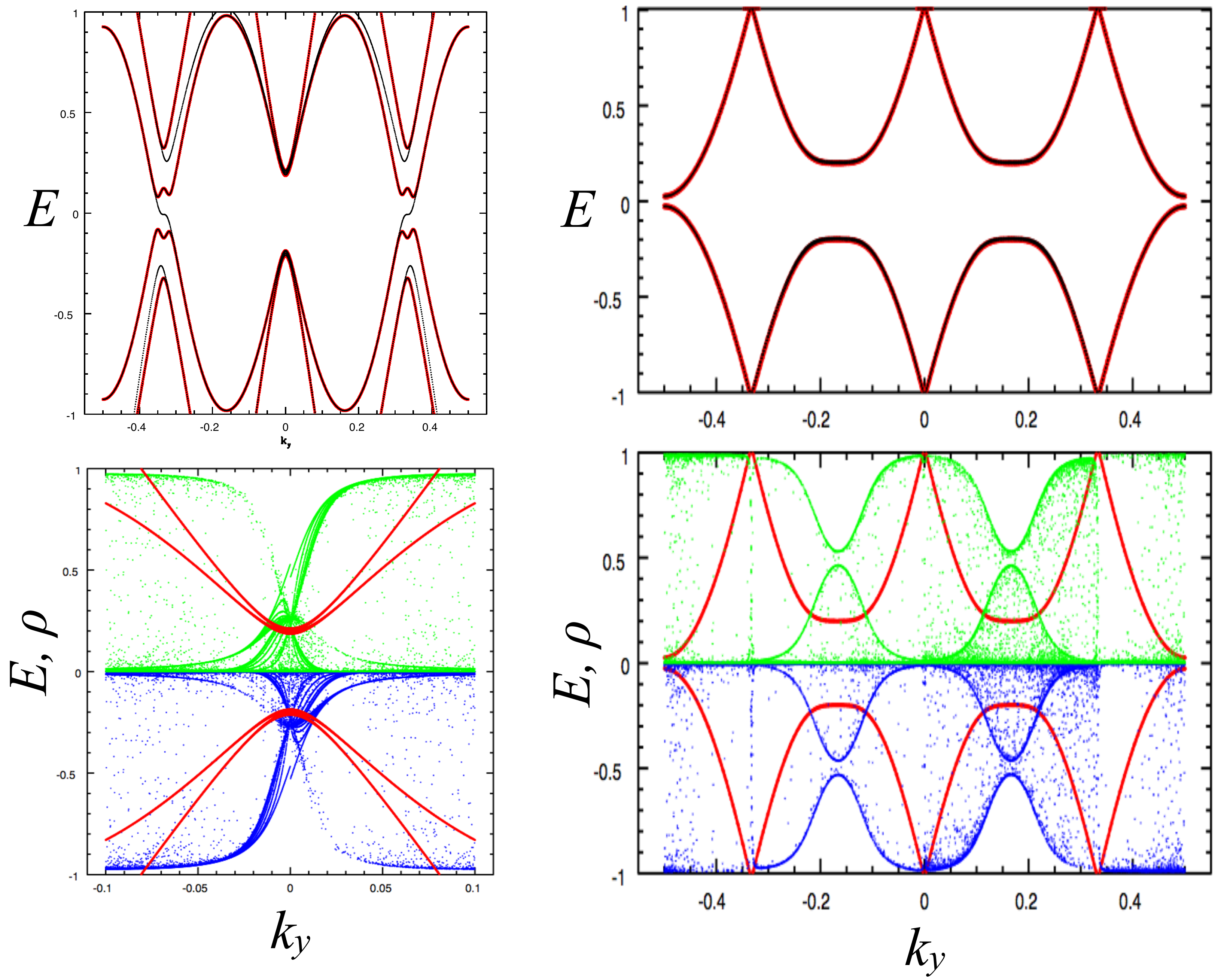}
\leavevmode \caption { Figure illustrates quantum phase transition from
topologically non-trivial state ( indicated by gapless edge modes ) to normal insulating state.
for $\gamma=.25$, $\lambda=5$ , and $\lambda_{stag}=5$ ( top-left) and $\lambda_{stag}=10$ (right).
In the bottom caption, Green and blue show the local density of the two-component state at all sites. For normalized
spinor wave function, a density of $.25$ shows the existence of a dimerized state for {\it each component} of the spinor. }
\label{qsh}
\end{figure}

For isotropic lattices, this Multiband system has been found to exhibit\cite{goldman} 
a series of quantum phase transitions between QSH, normal insulator and metallic phases in two-dimensional
parameter space for various values of chemical potential.
Figure ~(\ref{qsh})  
shows the landscape of the phases 
near half-filling for $\alpha=1/6$.
The transitions from topologically nontrivial quantum spin Hall (QSH) to topologically trivial normal band insulator  
is signaled by the disappearance of gapless Kramer-pair of chiral edge modes.
\\

Motivated by the fact that anisotropic lattices, reveal a novel type of encoding
of nontrivial topological states in the form of ``Chern-Dimers'', for $U(1)$ gauge systems, we present our preliminary results of QSH states for anisotropic lattices.
As shown in the Fig. ~\ref{qsh}, anisotropic lattices subjected to $SU(2)$ gauge fields
exhibit quantum phase transitions from QSH to normal insulating state by tuning the staggered potential.
In contrast to the QH system, the  avoided crossings in QSH system need not be isolated as seen
near $k_y=1/3$. However, there are states, such as those near $k_y=0$ where one sees an isolated avoided crossing.

Our preliminary investigation as summarized in figure ~(\ref{qsh}) shows that for anisotropic lattices,
in QSH phase, both components of the spinor
wave function are dimerized near an isolated avoided crossings. It is the existence of this double or 
spinor-dimer that distinguishes QSH state from a normal insulator.
In other words, each component of the wave function of topologically non-trivial insulating phase localizes at
two distinct sites, which is in contrast to the topologically trivial phase,
where the-components of the wave functions localize at a single site.
For normalized states, the existence of a double-dimer can be spotted by density approaching $.25$ as illustrated
in the Fig. \ref{qsh}. 
The distinction between the QSH and normal insulator,
characterized by spinor-dimer and its absence suggests that such topological phase transitions
may be identified by time of flight images.

\section*{Open questions and outlook}

Identifying topological states in time of flight images
would be a dream measurement for cold atom experimentalists. 
Therefore, our result that the momentum distribution carries fingerprints of nontrivial topology
is important.

Our studies raise several open questions regarding topological states and its manifestations.
Firstly, our detailed investigation 
suggests that the features of topological invariant (Chern number) in momentum distribution are quite robust.
Is there a rigorous theoretical framework that can establish this connection even for parameters beyond Chern-dimer regime?
Secondly, the sign of the Chern number can not be directly determined from the momentum distribution. 
It remains an open question whether some other observable can 
detect its sign.

Another natural question is: how generic is the existence of dimerized states in topological states of matter?
Somewhat analogous to edge states, the size of the dimerized states in our example is
a robust (topologically protected) quantity. Furthermore, are edge states and dimerized states
related to each other? Naively, one expects a dimer of size $N$ in bulk
will result in $N$-possible types of defects at open boundaries. These ``defects'' quite possibly will manifest as
edge states inside the gap. It is conceivable that the topological aspect of the dimers of size $N$ is intimately related
to $N$-edge states in QH systems. Further work is required to test these speculations,
even in special cases.

To investigate the generality of the dimer framework in topological states, we have
studied QSH states for anisotropic lattices.
Our preliminary studies suggest that the topological phase transition from QSH to normal
insulating state is accompanied by the absence of dimerized states and hence may be traceable
in time of flight images.
Detailed characterization of $Z_2$ invariance within the framework of spatially extended objects, however
remains appealing but open problem.

The subject of topological states of matter is an active frontier in physics.
Lessons learned from their realizations in ultracold atoms will 
help to advance the fundamental understanding, and inspire new material design or
technological applications. Ultracold topological superfluids, Mott insulators with 
nontrivial topology, and non-equilibrium topological states are most likely 
to bear fruit in reasonably near future.

We thank Noah Bray-Ali, Nathan Goldman, Ian Spielman, and Carl Williams for collaboration and their contributions during
the course of this research. We acknowledge the funding support from ONR and NIST.

\section*{Index}
Chern number....3,6,7 \\
Chern-Dimer mapping....6\\
Chern tree.... 11, 12, 13\\
Fibonacci numbers..... 12\\
Harper equation.....5\\
Hofstadter model....4\\
Incommensurate flux....11\\
Landau-level......9\\
\end{document}